\documentclass[preprint,tightenlines,aps,prd,showpacs,nofootinbib]{revtex4}
\usepackage{graphicx}
\usepackage{amsmath}
\usepackage{amssymb}
\usepackage{bm}
\begin{document}
\title{\mbox{}\\[10pt] Charm-pair Rescattering Mechanism \\ 
                       for Charmonium Production 
                       in High-energy Collisions}
\author{Pierre Artoisenet}
\affiliation{Center for Particle Physics and Phenomenology,
Universit\'e Catholique de Louvain, B1348 Louvain-la-Neuve, Belgium}
\author{Eric Braaten}
\affiliation{Physics Department, Ohio State University, Columbus, OH
  43210, USA}
\affiliation{Bethe Center for Theoretical Physics,
	Universit\"at Bonn, 53115 Bonn, Germany}
\date{\today}
\begin{abstract}
A new mechanism for heavy quarkonium production in high-energy collisions 
called the ``$s$--channel cut'' was proposed in 2005 by Lansberg,
Cudell, and Kalinovsky.  We identify this mechanism physically as
the production 
of a heavy quark and anti-quark that are on-shell followed by
their rescattering to produce heavy quarkonium.
We point out that in the NRQCD factorization formalism 
this rescattering mechanism is a contribution 
to the color-singlet model term at 
next-to-next-to-leading order in perturbation theory. 
Its leading contribution to the production rate can be calculated without 
introducing any additional phenomenological parameters.
We calculate the charm-pair rescattering (or $s$--channel cut)
contribution to the production 
of $J/\psi$ at the Tevatron and compare it to
estimates by Lansberg et al.\ using phenomenological models.
This contribution competes with the leading-order term
in the color-singlet model  at large transverse momentum but is 
significantly smaller than the next-to-leading-order term.
We conclude that charm-pair rescattering is not a dominant mechanism 
for charmonium production in high-energy collisions. 
\end{abstract}



\maketitle

\section{Introduction}

The production of heavy quarkonium in high energy collisions remains a
challenging problem in QCD.  This problem has some aspects that are 
inherently perturbative, because it requires the creation of a heavy
quark-antiquark pair. It also has aspects that are inherently nonperturbative,
because the heavy quark $Q$ and the antiquark $\bar Q$ are required to
form a bound state. The nonrelativistic QCD (NRQCD) factorization approach 
uses an effective field theory to systematically separate
perturbative short-distance aspects from nonperturbative
long-distance aspects~\cite{Bodwin:1994jh}. 
The inclusive production cross section is
expressed as the sum of products of parton scattering factors that can be
calculated using perturbative QCD and NRQCD matrix elements that can be
treated as phenomenological parameters. The NRQCD factorization formulas
have not yet been proven rigorously, but they are well-motivated by
proofs of factorization to all orders of perturbation theory for other 
processes in high energy physics.  The NRQCD factorization formula
for gluon fragmentation into quarkonium has
survived a nontrivial test at 
next-to-next-to-leading order, although it 
required a modification of the formal definition of the
NRQCD matrix elements~\cite{Nayak:2005rt}.

NRQCD predicts a definite hierarchy in the size of the NRQCD matrix
elements according to how they scale with the typical relative velocity
$v$ of the heavy quark in quarkonium~\cite{Bodwin:1994jh}. However, the NRQCD
factorization framework is sufficiently flexible that any model for
quarkonium production that is compatible with perturbative QCD at short
distances should be expressible in terms of assumptions about the NRQCD
matrix elements. The color-singlet model (CSM) can be expressed as the
vanishing of all NRQCD matrix elements except one. The CSM matrix element
corresponds to formation of the quarkonium from $Q \bar Q$ in a
color-singlet state with the same angular-momentum quantum numbers as the
quarkonium in the potential model. 
The color-evaporation model can be expressed as the NRQCD matrix elements
having a hierarchy according to the orbital angular momentum of the 
$Q \bar Q$ pair without regard to their color 
and spin quantum numbers~\cite{Bodwin:2005hm}.

In 2005, Lansberg, Cudell, and Kalinovsky (LCK) proposed a new mechanism
for heavy quarkonium production that they called the 
``$s$--channel'' cut~\cite{Lansberg:2005pc}.
In the case of the production of quarkonium from gluon-gluon
collisions, they claimed it was leading order in the QCD coupling
constant but had not been taken into account in any previous analyses.
The name ``$s$--channel cut'' refers to a mathematical description of the
mechanism. It can be identified as a contribution to the imaginary part
of the amplitude for quarkonium production in gluon-gluon collisions that
corresponds to a Cutkosky cut through $Q$ and $\bar Q$ lines only. It is
thus associated with a discontinuity of the amplitude in the square 
$s$ of the center-of-mass energy of the colliding gluons. 
This mechanism corresponds physically
to the creation of the heavy quark and antiquark through the parton
scattering process $gg \to Q \bar Q$ followed by the formation of
quarkonium through the rescattering of the $Q \bar Q$ pair.
A physical description of this mechanism is $Q \bar Q$ 
rescattering into quarkonium plus a recoiling gluon.

LCK applied their $s$--channel cut mechanism to $J/\psi$ production in
hadron-hadron collisions. They expressed the $s$--channel cut contribution
to the production amplitude as the sum of diagrams with a
vertex $\Gamma_\mu$ for $c \bar c \to J/\psi$ and diagrams with 
another vertex $\Gamma_{\mu \nu}$ for $c \bar c \to J/\psi \, g$ 
that was introduced in order to
restore a Ward identity. They developed simple phenomenological models
for these vertices and used them to estimate the $s$--channel cut
contribution to the inclusive $J/\psi$ cross section in $p \bar p$ and
$pp$ collisions~\cite{Lansberg:2005pc,Haberzettl:2007kj}.

Since the $Q \bar Q$ rescattering (or $s$--channel cut) mechanism involves
the formation of quarkonium from a $Q \bar Q$ pair that is created by a
perturbative parton process, it must be expressible in terms of the
conventional NRQCD factorization formula. In this paper, we identify the
charm-pair rescattering contribution to the process $gg \to J/\psi \, g$
as a next-to-next-to-leading order (NNLO) contribution to the
color-singlet model term in the NRQCD factorization formula. It can
therefore be calculated without introducing any new phenomenological
parameters.

In Section~\ref{sec:s-cut}, we summarize the results of Lansberg et al.\ 
on the $s$--channel cut mechanism for $J/\psi$ production in hadron-hadron
collisions. In Section~\ref{sec:Charm-rescat}, we show that this mechanism, 
which can be described physically as charm-pair rescattering, 
is an NNLO contribution to the CSM term in the NRQCD factorization formula. 
In Section~\ref{sec:NRQCD}, we explain how the charm-pair rescattering 
contribution can be calculated without introducing any new 
phenomenological parameters.  In Section~\ref{sec:calculation}, 
we compare our numerical results for $J/\psi$ production 
at the Tevatron with those 
from the phenomenological models introduced by LCK~\cite{Lansberg:2005pc} 
and by Lansberg and Haberzettl (LH) in Ref.~\cite{Haberzettl:2007kj}. We also
compare our results to other contributions to the CSM term at leading
order (LO) and next-to-leading order (NLO) in the QCD coupling constant.

\section{$\bm{s}$--channel cut}
\label{sec:s-cut}

\begin{figure}[t]
\center
\includegraphics[scale=0.9]{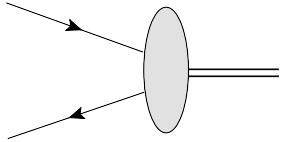} 
\caption{Vertex $\Gamma_\mu$ for $c$ and $\bar c$ to form $J/\psi$.} 
\label{fig1}
\end{figure}
\begin{figure}[t]
\center
\includegraphics[scale=0.9]{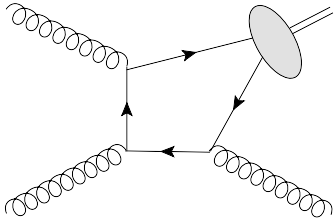} 
\caption{One of the 6 Feynman diagrams for $g \, g \to J/\psi \, g$ 
at order $g_s^3$ with vertex $\Gamma_\mu$.} 
\label{fig2}
\end{figure}

We begin by summarizing the $s$--channel cut mechanism for charmonium 
production proposed by Lansberg, Cudell, and Kalinovsky 
(LCK)~\cite{Lansberg:2005pc}. 
Their starting point was the 
amplitude $\Gamma_\mu (q,P) \epsilon^\mu$ for a color-singlet $c \bar c$ 
pair with total 4-momentum $P$ and relative 4-momentum $q$ to form 
$J/\psi$ with 4-momentum $P$ and polarization vector $\epsilon$. 
This  vertex is illustrated in Fig.~\ref{fig1}. 
It can be defined diagrammatically by starting with the sum of
all diagrams with two incoming $c$ and $\bar c$ lines and two outgoing 
$c$ and $\bar c$ lines.  This sum defines an amplitude with a pole 
in the $c \bar c$ invariant mass at the mass of the $J/\psi$. 
In the residue of the pole, the desired amplitude 
$\Gamma_\mu \epsilon^\mu$ is the factor associated with the 
incoming $c$ and $\bar c$.  Any amplitude for the production of 
$J/\psi$ can be expressed as a sum of diagrams 
with a vertex $\Gamma_\mu$ and additional QCD vertices. 
For example, the amplitude for $g \, g \to J/\psi \, g$ at order $g_s^3$ has 
contributions from 6 one-loop box diagrams, 
one of which is shown in Fig.~\ref{fig2}.
The other five diagrams are obtained by permuting the attachments 
of the three gluons.  For general quarkonium states, 
there are also one-loop triangle diagrams involving a 3-gluon vertex.
By setting the incoming and outgoing gluons on their mass shells, 
we obtain the T-matrix element for this process.

\begin{figure}[t]
\center
\includegraphics[scale=0.9]{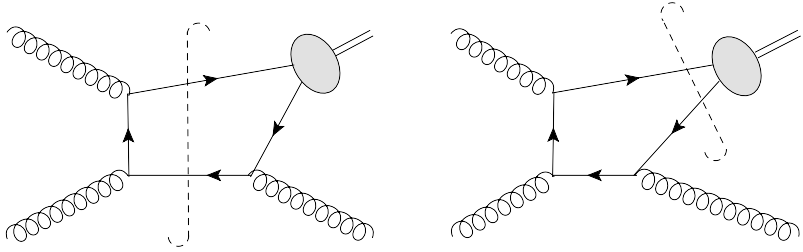} 
\caption{The $s$--channel $c \bar c$ cut (a) and the meson $c \bar c$ cut
(b) for the Feynman diagram in Fig.~\ref{fig2}.} 
\label{fig3}
\end{figure}

A Feynman diagram has a discontinuity in the invariant mass 
of any set of lines that are kinematically allowed to be on shell. 
This discontinuity can be represented by a cut diagram 
in which the cut passes through that set of lines.
There are two possible cuts through the parton lines 
in the diagram in Fig.~\ref{fig2}:
\begin{itemize}

\item  
the {\it meson $c \bar c$ cut}, in which the $c$ and $\bar c$ lines 
that attach to the $\Gamma_\mu$ vertex are cut, 
as illustrated in Fig.~\ref{fig3}(b). 
The discontinuity in the amplitude is in the invariant mass $P^2$ 
of the $c \bar c$ pair. 
The cut diagram is the product of diagrams for $g \, g \to c \, \bar c \, g$ and 
$c  \, \bar c \to J/\psi$ integrated over the relative momentum 
of the $c$ and $\bar c$, which are on shell.  All 6 diagrams of order $g_s^3$
have a meson cut if the charm quark mass 
satisfies $m_c < M_{J/\psi}/2$.

\item  
the {\it $s$--channel $c \bar c$ cut}, in which the $c$ and $\bar c$ lines 
that are attached to the incoming gluons but do not join them are cut, 
as illustrated in Fig.~\ref{fig3}(a). The discontinuity in the amplitude
is in the invariant mass $s$ of the colliding gluons. 
The cut diagram is the product of diagrams for  
$g \, g \to c \, \bar c$ and  $c \, \bar c \to J/\psi \, g$ 
integrated over the relative momentum of the $c$ and $\bar c$, 
which are on shell.
Four of the 6 diagrams of order $g_s^3$
have an $s$--channel cut if $s > M_{J/\psi}^2$.

\end{itemize}

In the color-singlet model for quarkonium production, 
the production amplitude for $J/\psi$ is a product of a 
wavefunction factor and the 
amplitude for producing a $c$ and $\bar c$ that are on their
mass shell in a color-singlet 
$^3S_1$ state with relative momentum 0 and total momentum equal 
to that of the $J/\psi$. 
In an amplitude for $J/\psi$ production with vertex $\Gamma_\mu$,
the meson $c \bar c$ cut puts the $c$ and $\bar c$ on shell and therefore 
gives a contribution that resembles the amplitude in the 
color-singlet model.
In the amplitude for $g \, g \to J/\psi \, g$ at order $g_s^3$
with vertex $\Gamma_\mu$, LCK identified
the meson cut with the amplitude in the color-singlet model 
and they argued that the $s$--channel cut gives an additional 
leading-order contribution that had not previously been taken into account.
As will be explained in Section~\ref{sec:Charm-rescat},
the meson $c \bar c$ cut is subleading compared to the conventional 
color-singlet model amplitude 
and the $s$--channel cut is actually a contribution to the
color-singlet model amplitude at next-to-next-to-leading 
order in the running coupling constant of QCD.

We now describe the formalism that LCK used to calculate the 
$s$--channel cut contribution to $J/\psi$ production. They assumed that 
the amplitude $\Gamma_\mu$ was dominated by the contribution from the 
$c \bar c$ Fock state and has the form $\Gamma (q_{\rm rel}) \gamma_\mu$, 
where $\Gamma$ is a scalar function of the relative 3-momentum 
$q_{\rm rel}$ of $c$ and $\bar c$ in the $J/\psi$ rest frame. 
This reduces $\Gamma_\mu$ to a single function of one variable. 
They parameterized $\Gamma_\mu$ as a Gaussian with 
normalization factor $N$ and a momentum scale $\Lambda$. 
This reduces $\Gamma_\mu$ to two phenomenological parameters. 
They used results from relativistic potential models to determine 
$\Lambda$ to be approximately 1.8~GeV
and they used the electronic width of $J/\psi$ to determine $N$. 

\begin{figure}[t]
\center
\includegraphics[scale=0.9]{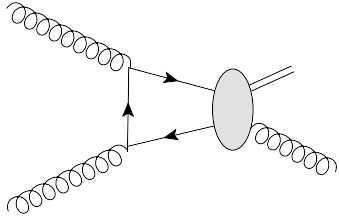} 
\caption{One of the 2 Feynman diagrams for $g \, g \to J/\psi \, g$ 
at order $g_s^2$ with vertex $\Gamma_{\mu \nu}$ in the model 
for the $s$--channel cut contribution introduced in
Ref.~\cite{Lansberg:2005pc}.} 
\label{fig4}
\end{figure}

LCK recognized that the vertex $\Gamma_\mu$ is not gauge-invariant
when the incoming $c$ and $\bar c$ are not on their mass shells.
One consequence is that the $s$--channel cut to the amplitude for 
$g \, g \to J/\psi \, g$ at order $g_s^3$ with vertex $\Gamma_\mu$
does not satisfy the Ward identity 
associated with the current that creates the outgoing gluon.
They proposed that the Ward identity 
could be restored by adding diagrams with a 4-point vertex
$\Gamma_{\mu \nu} (p_1, p_2, q, P)$ for $c \bar c \to J/\psi \, g$.
They did not give any diagrammatic definition of this vertex,
but it was assumed to be of order $g_s$ relative to $\Gamma_\mu$.
There are two diagrams for $g \, g \to J/\psi \, g$ 
at order $g_s^2$ with a vertex $\Gamma_{\mu \nu}$, both of which have an  
$s$--channel cut.  One of the diagrams is shown in Fig.~\ref{fig4} 
and the other is obtained by reversing the arrows on the $c$ and $\bar c$ lines. 

The complete model of LCK for the $s$--channel cut contribution to the 
T-matrix element for $g \, g \to J/\psi \, g$ at order $g_s^3$
is the sum of the 4 cut diagrams with vertex $\Gamma_\mu$
and the 2 cut diagrams with vertex $\Gamma_{\mu \nu}$.
They took the charm quark mass to be $m_c=1.87$~GeV, which is
greater than $M_{J/\psi}/2$ so there is no 
meson $c \bar c$ cut.  In Ref.~\cite{Lansberg:2005pc}, 
LCK proposed an expression for the vertex $\Gamma_{\mu \nu}$ 
with no additional parameters that respects the Ward identity.
Thus this model for the $s$--channel cut contribution has 2 parameters, 
$N$ and $\Lambda$, in addition to $m_c$ 
and the QCD coupling constant $\alpha_s$.
LCK showed that in this model the cross section for $J/\psi$ production 
at the Tevatron at large transverse momentum ($p_T$) 
could be as large as the leading-order color-singlet model contribution.
In Ref.~\cite{Haberzettl:2007kj}, 
LH proposed a more general expression for $\Gamma_{\mu \nu}$ that depends on a 
phenomenological function.  They introduced a simple parameterization 
for this function that depends on two parameters.
Thus this model for the $s$--channel cut contribution has 4 parameters
in addition to $m_c$ and $\alpha_s$.
LH showed that if these parameters are adjusted so that
the $s$--channel cut contribution alone fits the 
Tevatron data for $J/\psi$ production at low $p_T$, 
the $s$--channel cut contribution alone also 
fits the RHIC data on $J/\psi$ production.

\section{Charm-pair rescattering}
\label{sec:Charm-rescat}

As pointed out in the Introduction, the $s$--channel $c \bar c$ 
cut mechanism for $J/\psi$ production can be interpreted physically 
as charm-pair rescattering.
The $s$--channel cut to the one-loop amplitude  
with vertex $\Gamma_\mu$ for $g \, g \to J/\psi \, g$ 
separates the incoming gluons from the outgoing $J/\psi$
and gluon. By the Cutkosky cutting rules, the imaginary part 
generated by the discontinuity of the amplitude in the variable $s$
 is the sum over all such cuts.
All the other cuts that contribute to that imaginary part must
cut through the vertex $\Gamma_\mu$. 
(The meson $c  \bar c$ cut is not one of these cuts,
because it does not separate the incoming lines from the outgoing lines.)
A Cutkosky cut through a set of parton lines corresponds to the scattering 
of $g \, g$ into this set of on-shell partons followed by the 
rescattering of those partons into
$J/\psi \, g$. The $s$--channel cut discussed by LCK is through $c$ and $\bar c$
lines, so it corresponds to the scattering of $g \, g$ into
$c  \, \bar c$ followed by the rescattering of $c  \, \bar c$ into $J/\psi \, g$.

One misconception in Ref.~\cite{Lansberg:2005pc} (LCK) is that the
meson $c \bar c$  cut of the amplitude for $J/\psi$ production should be
identified with the color-singlet model contribution.
At first glance, this identification may seem plausible, 
because the meson cut puts the $c$ and $\bar c$ on their mass shells 
and the color-singlet model amplitude is usually calculated 
by taking the $c$ and $\bar c$ lines 
to be on their mass shells with 0 relative momentum.
However this identification implies incorrectly that if the 
arbitrary phase of $\Gamma_\mu$ is chosen so that its dominant 
contribution is real, the leading color-singlet model contribution 
is pure imaginary.  The color-singlet model amplitude should instead be
identified with the contribution from the region of phase space 
that gives the dominant contributions to $\Gamma_\mu$.
Since the $J/\psi$ is a nonrelativistic bound state, 
its primary constituents $c$ and $\bar c$ 
have typical relative velocity $v$  that is significantly smaller than 1.
The largest contributions to $\Gamma_\mu$ come from the region in 
which the $c$ and $\bar c$ have relative momentum of
order $v$ in the $J/\psi$ rest frame and energies that differ by $m_c$ by
amounts of order $m_cv^2$.  Calculations in the color-singlet model
in which $c$ and $\bar c$ are on their mass shells with 0 relative 
momentum should be understood simply as a technical device for 
calculating the contributions from this dominant region.
If the arbitrary phase of $\Gamma_\mu$ is chosen so that its dominant 
contribution is real, the leading color-singlet model contribution 
will also be real.  

By presenting the s-channel $c \bar c$ cut as a leading order 
contribution, LCK implicitly assumed that the 
vertex $\Gamma_\mu$ for a $c \bar c$ pair to form $J/\psi$ can be treated
as if it has a definite order in the QCD coupling constant $g_s$. However
its order in $g_s$ actually depends on the relative momentum of the 
$c \bar c$ pair. The running coupling constant $g_s(\mu)$ is small at momentum
scales $\mu$ of order $m_c$ and becomes large at small momentum scales. 
It is convenient to treat $g_s(\mu)$ as being of order 1 at 
small momentum scales $\mu \ll m_c$. 
We then need only count powers of the small running
coupling constant $g_s(m_c)$ from QCD vertices with momentum scale $\mu$ of
order $m_c$ and larger. If the relative momentum of the $c$ and $\bar c$ 
is small compared to $m_c$, then $\Gamma_\mu$ is of order $g_s(m_c)^0 = 1$. 
We will refer to the vertex $\Gamma_\mu$ in this momentum region as a
``soft vertex". If the relative momentum of $c$ and $\bar c$ is of order
$m_c$, large momentum must be transferred between the $c$ and $\bar c$ by
the exchange of hard gluons with coupling constant $g_s(m_c)$
in order for the $c$ and $\bar c$ to bind to form $J/\psi$.  The lowest order
contribution comes from one-gluon exchange, so $\Gamma_\mu$ is of order
$g_s(m_c)^2$. We will refer to $\Gamma_\mu$ in this momentum region as a
``hard vertex".

\begin{figure}[t]
\center
\includegraphics[scale=0.9]{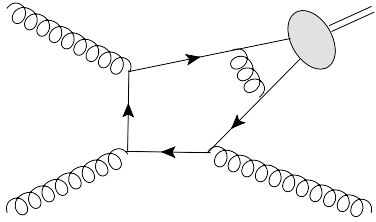} 
\caption{Two-loop Feynman diagrams for $g \, g \to J/\psi \, g$ 
with a soft vertex $\Gamma_\mu$.  This diagram of order $g_s(m_c)^5$ 
is also a contribution to the diagram in Fig.~\ref{fig2} 
with a hard vertex $\Gamma_\mu$.} 
\label{fig5}
\end{figure}

\begin{figure}
\center
\includegraphics[scale=1]{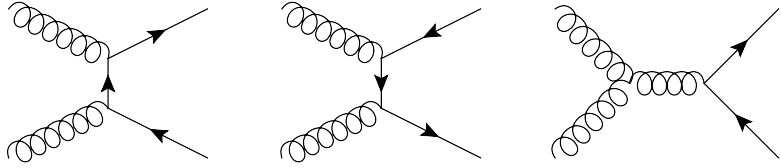}
\caption{Feynman diagrams for  $g \, g \rightarrow c \, \bar c $.}
\label{FD_gg_cc}
\end{figure}

\begin{figure}
\center
\includegraphics[scale=1]{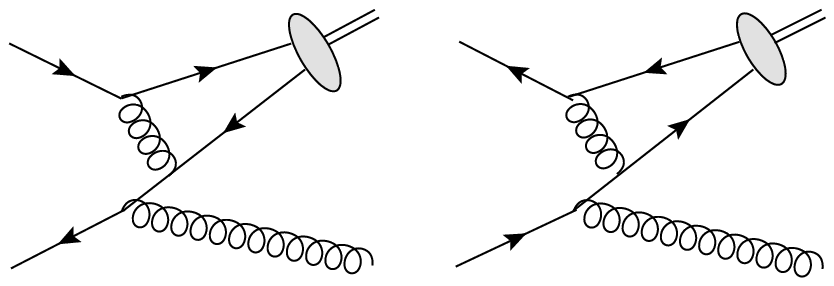}
\caption{Two of the 5 one-loop Feynman diagrams for  
$c \, \bar c \rightarrow J/\psi \, g$
with a soft-vertex $\Gamma_\mu$.}
\label{FD_cc_gcc}
\end{figure}

\begin{figure}
\center
\includegraphics[scale=1]{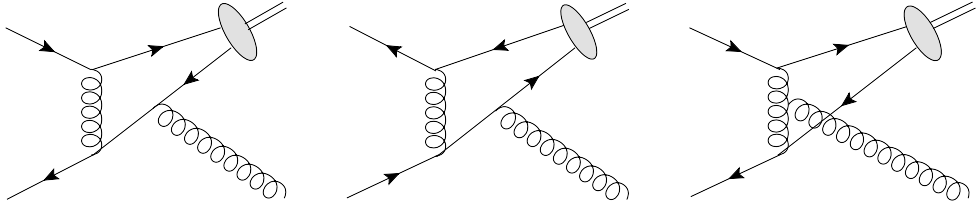}
\caption{Three of the 5 one-loop Feynman diagrams for  
$c \, \bar c \rightarrow J/\psi \, g$
with a soft-vertex $\Gamma_\mu$.}
\label{FD_cc_gccB}
\end{figure}

In a one-loop diagram for $g \, g \to J/\psi \, g$ with vertex $\Gamma_\mu$, 
such as the diagram in Fig.~\ref{fig2}, 
the $s$--channel cut forces either the $c$ or
$\bar c$ that enters the vertex to be off-shell by an amount
of order $m_c$ or larger. Thus $\Gamma_\mu$ is a hard vertex of order
$g_s^2(m_c)$ and the diagram is of order $g_s(m_c)^5$. 
The leading contribution to the hard vertex comes from the exchange 
of a single hard gluon, which can be pulled out of the hard vertex as
illustrated in Fig.~\ref{fig5}, changing it into a soft vertex.
One-loop diagrams with a hard vertex, like the one in Fig.~\ref{fig2}, 
can therefore be expressed as two-loop diagrams with a soft vertex, 
like the one in Fig.~\ref{fig5}.
The cancellations that
restore the Ward identity must come from other diagrams of order
$g_s(m_c)^5$.  Since the complete amplitude for $J/\psi$ production can be
expressed as a sum of diagrams with a soft vertex $\Gamma_\mu$, the
cancellations must come specifically from other
two-loop diagrams with a soft vertex $\Gamma_\mu$. 
There are hundreds of two-loop diagrams with a vertex $\Gamma_\mu$,
but only 15 of them have an $s$--channel $c \bar c$ cut.
They can be obtained by connecting the $c$ lines and the $\bar c$ lines in 
the products of the three diagrams for $g \, g \rightarrow c \, \bar c $
in Fig.~\ref{FD_gg_cc} and the five diagrams for 
$c \, \bar c \rightarrow c \bar c_1(^3S_1) \, g$ 
in Figs.~\ref{FD_cc_gcc} and \ref{FD_cc_gccB}.

If we retain only this leading s-channel $c \bar c$ cut contribution,
the phenomenological model of LCK
has a straightforward interpretation in terms of 
the diagrams in Figs.~\ref{FD_gg_cc}, \ref{FD_cc_gcc}, and \ref{FD_cc_gccB}.
The 4 cut diagrams of LCK with vertex $\Gamma_\mu$ can be identified 
with the products of the first two diagrams for $gg \to c \bar c$ 
in Fig.~\ref{FD_gg_cc} and the two diagrams for 
$c \, \bar c \rightarrow J/\psi \, g$ in Fig.~\ref{FD_cc_gcc}.
The third diagram for $gg \to c \bar c$ 
in Fig.~\ref{FD_gg_cc}, which involves a 3-gluon vertex, does 
not contribute at leading order because of the 
$^3S_1$ quantum numbers of the $J/\psi$.
The vertex $\Gamma_\mu$ used by LCK is a phenomenological model 
for the hard vertex obtained by absorbing the exchanged gluon into 
the soft vertex in the two diagrams of Fig.~\ref{FD_cc_gcc}.
The 2 cut diagrams of LCK with vertex $\Gamma_{\mu \nu}$ can be identified 
with the products of the first two diagrams for $gg \to c \bar c$ 
in Fig.~\ref{FD_gg_cc} and the 3 diagrams for 
$c \, \bar c \rightarrow J/\psi \, g$ in Fig.~\ref{FD_cc_gccB}.
Thus, at leading order, LCK's vertex $\Gamma_{\mu \nu}$ is just an ad-hoc model 
for the sum of the 3 diagrams in Fig.~\ref{FD_cc_gccB}.


This argument shows that the s-channel $c \bar c$ 
cut is a contribution to the color-singlet amplitude 
that is subleading in the strong coupling constant.
Indeed, the leading contribution to the amplitude for 
$g \, g \to J/\psi \, g$ for a $J/\psi$ with transverse momentum 
of order $m_c$ comes from one-loop diagrams with a soft vertex 
$\Gamma_\mu$, like the one in Fig.~\ref{fig2}.
The amplitude is therefore of order $g_s^3(m_c)$, 
so the cross section is of order $\alpha_s^3(m_c)$.   
With the appropriate choice of the phase of $\Gamma_\mu$,
the leading contribution to the amplitude is real-valued 
and can be identified with the CSM amplitude.
The charm-pair rescattering amplitude 
from the $s$--channel cut for 
$g \, g \to J/\psi  \, g$ is of order $g_s^5(m_c)$
and is pure imaginary.
There is no interference between this charm-pair rescattering amplitude 
and the leading term in the CSM amplitude. 
The cross section from charm-pair rescattering is therefore of order
$\alpha_s^5(m_c)$, which is NNLO in the QCD coupling constant.

\section{NRQCD calculation}
\label{sec:NRQCD}

We now consider the inclusive cross section for $J/\psi$ production in 
$p \bar p$ collisions. At leading order in $\alpha_s$, the only parton
process that contributes to the CSM term is 
$gg \to c \bar c_1(^3S_1) \, g$, where $c \bar c_1(^3S_1)$ represents 
a $c$ and $\bar c$ in a color-singlet spin-triplet state 
whose momenta are both equal to half that of the $J/\psi$. 
The contribution of this parton process to the cross section is
\begin{equation}
\sigma_{\rm CSM}[p \bar p \rightarrow J/\psi + X] =
\int dx_1 dx_2 f_{g/p}(x_1) f_{g/\bar p}(x_2) 
\hat \sigma[g \, g \rightarrow  c \bar c_1(^3S_1) \, g]
\langle \mathcal O_1^{J/\psi}(^3S_1) \rangle .
\label{total_cross_section}
\end{equation}
The parton cross section $\hat \sigma$ can be expressed as an
integral over the solid angle $\Omega_{c\bar c}$ for the
momentum of the $c \bar c$ pair:
\begin{equation}
\hat \sigma
\left[ g \, g \rightarrow c \bar c_1(^3S_1) \, g \right]
= \frac{\hat s - 4 m_c^2}{64\pi^2 \hat s^2} \int d\Omega_{c \bar c}
\left| \mathcal{M} [g \, g \rightarrow c \bar c_1(^3S_1) \, g] \right|^2.
\label{SD_coefficient}
\end{equation}
The prefactor comes from the 2-body phase space for a 
$c \bar c$ pair and a gluon multiplied by the flux factor 
$1/(2 \hat s)$, where $\hat s$ is the square of the 
invariant mass of the colliding gluons (which was denoted by $s$ 
in the previous sections). The leading order
term in the invariant matrix element $\mathcal{M}$ 
comes from tree diagrams, such as the
diagram in Fig.~\ref{fig2} with the blob omitted, and is order $g_s^3$. 
This term is real-valued provided a real-valued basis 
for the polarization vector of the gluons is used. 
The leading imaginary terms in the invariant matrix element 
are of order $g_s^5$. Using the Cutkosky cutting rules, 
the imaginary part can be expressed as a sum over the $s$--channel cuts 
through the diagrams, which separate the incoming gluons from the
$c \bar c$ pair and the gluon in the final state. At order $g_s^5$, the
only $s$--channel cuts are $gg$ cuts and $c \bar c$ cuts. 
The contribution from the $s$--channel $c \bar c$ cut, which
is the charm-pair rescattering term, can be expressed as in integral over
the solid angle $\Omega_c$ of the charm quark in the intermediate state:
\begin{equation}
{\rm Im} \mathcal{M}_{c \bar c \ {\rm cut}}
	[g \, g \rightarrow c \bar c_1(^3S_1) \, g] =
 \frac{(\hat s - 4 m_c^2)^{1/2}}{64 \pi^2 \hat s^{1/2}} 
\int d \Omega_c \sum
\mathcal{M}[g \, g \rightarrow c \, \bar c]
\mathcal{M}[ c \, \bar c \rightarrow c \bar c_1(^3S_1) \, g].
\label{amp_gg_cc_ccg}
\end{equation}
The prefactor comes from multiplying the 2-body phase space for $c$ and
$\bar c$ by a factor of 1/2 from the optical theorem. The sum is over the
helicities and the color indices of the intermediate $c$ and $\bar c$.
At leading order, the first invariant matrix element $\mathcal{M}$ 
on the right side of Eq.~(\ref{amp_gg_cc_ccg}) 
is the sum of the three Feynman diagrams for $g \, g \to c \, \bar c$ in 
Fig.~\ref{FD_gg_cc}, whereas the second invariant matrix element $\mathcal{M}$ 
on the right side of Eq.~(\ref{amp_gg_cc_ccg}) is the sum of the 
five Feynman diagrams for
$c \, \bar c \to c \bar c_1(^3S_1) \, g$ in Figs.~\ref{FD_cc_gcc}
and \ref{FD_cc_gccB} with the blobs omitted. 
In the expression for Im${\mathcal{M}}_{c \bar c \, {\rm cut}}$
that comes directly from the optical theorem, the last factor in 
Eq.~(\ref{amp_gg_cc_ccg}) is the complex conjugate of the 
invariant matrix element for $c \bar c_1(^3S_1) \, g \to c \, \bar c$. 
The time reversal symmetry of QCD has been used
to express this as the invariant matrix element for 
$c \, \bar c \to c \bar c_1(^3S_1) \, g$.

The charm-pair rescattering contribution to the CSM cross section 
is obtained by replacing the invariant matrix element $\mathcal{M}$
in Eq.~(\ref{SD_coefficient}) by the contribution to its imaginary part
from the $c \bar c$ cut given in Eq.~(\ref{amp_gg_cc_ccg}),
and then inserting the resulting expression for the parton cross section 
$\hat \sigma$ into the $J/\psi$ production cross section 
in Eq.~(\ref{total_cross_section}).
We proceed to explain how we calculate this cross section numerically. 
We use a helicity basis for the polarization vectors of the gluons 
and the $c \bar c$ pair. The projection of the matrix element for 
$g \, g \to c \bar c \, g$ onto the appropriate $c \bar c$ helicity state 
is described in Ref.~\cite{Artoisenet:2007qm}. 
We use MadGraph~\cite{Alwall:2007st} to evaluate the helicity amplitudes 
$\mathcal{M}[g \, g \rightarrow c \, \bar c]$ and
$\mathcal{M}[ c \, \bar c \rightarrow c \bar c_1(^3S_1) \, g]$ 
numerically. The sum over the color indices of the
intermediate $c \bar c$ pair is carried out by calculating the color
factors by hand. The replacement of $\mathcal{M}$ in Eq.~(\ref{SD_coefficient}) 
by the expression for Im$\mathcal{M}$ in Eq.~(\ref{amp_gg_cc_ccg}) 
gives a 3-fold angular integral.  Together
with the integration over parton momentum fractions in 
Eq.~(\ref{total_cross_section}), we have
an 8-dimensional integral that is calculated numerically using the
adaptive integration program VEGAS~\cite{Lepage:1980dq}.

\section{Numerical results}
\label{sec:calculation}

In this Section, we calculate the charm-pair rescattering contribution 
to the CSM term in the inclusive cross section for 
direct $J/\psi$ production in $p \bar p$ collisions at the Tevatron.
We compare our results, which have no additional parameters,
to those from models for the $s$--channel cut contributions proposed by 
Lansberg et al.~\cite{Lansberg:2005pc,Haberzettl:2007kj}.
We also compare our charm-pair rescattering contribution,
which is a contribution to the CSM term at NNLO, 
to the CSM terms at LO and NLO~\cite{Campbell:2007ws,Gong:2008sn}.

\begin{figure}[t]
\center
\includegraphics[scale=0.9]{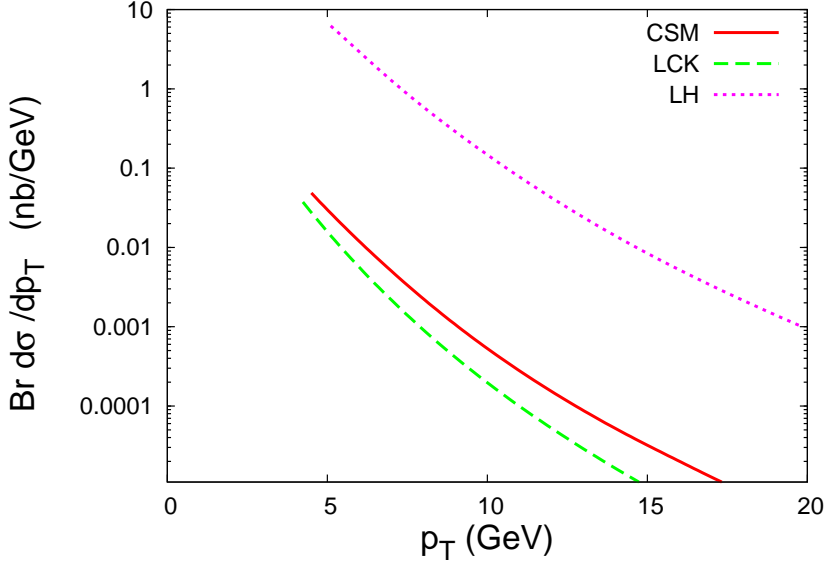} 
\caption{Differential cross section for direct $J/\psi$ 
production at the Tevatron from charm-pair rescattering
($s$--channel cut).
The differential cross section $d \sigma/dp_T$ 
integrated over the pseudorapidity interval $|\eta| < 0.6$ and
multiplied by the branching fraction for
$J/\psi \rightarrow \mu^+ \mu^-$ is shown as a function of $p_T$.
The curves are the CSM result (solid line), the result
from the model of Ref.~\cite{Lansberg:2005pc} (dashed line), and 
the result from the model of Ref.~\cite{Haberzettl:2007kj} (dotted line).
} 
\label{comp_predictions}
\end{figure}

The parton cross section for charm-pair rescattering in the 
color-singlet model depends only on $\alpha_s$, $m_c$, 
and the NRQCD matrix element $\langle \mathcal O_1^{J/\psi}(^3S_1) \rangle$.
To compare our results with those of Lansberg et al., 
we use parameters that are as close to theirs as possible.
They used a large value of the charm quark mass:
$m_c = 1.87$~GeV.  They determined the normalization factor 
in their vertex $\Gamma_\mu$ by fitting the partial width 
for $J/\psi \to e^+ e^-$.  The corresponding factor in the 
color-singlet model is $\langle \mathcal O_1^{J/\psi}(^3S_1) \rangle$.
At leading order in $\alpha_s$, the NRQCD factorization formula 
for the partial width for $J/\psi \to e^+ e^-$ is 
\begin{equation}
\Gamma[J/\psi \to e^+ e^-] = \frac{8 \pi \alpha^2}{81 m_c^2}
\langle \mathcal O_1^{J/\psi}(^3S_1) \rangle ~,
\end{equation}
where we have used the relation
$\langle \mathcal O_1^{J/\psi}(^3S_1) \rangle
   = 3 \langle \mathcal O_1(^3S_1) \rangle_{J/\psi}$
between the standard NRQCD production 
and decay matrix elements~\cite{Bodwin:1994jh}.  
Setting $m_c = 1.87$~GeV and $\alpha = 1/129.6$, 
we determine the NRQCD matrix element to be
$\langle \mathcal O_1^{J/\psi}(^3S_1) \rangle = 1.05$~GeV$^3$.
In Refs.~\cite{Lansberg:2005pc,Haberzettl:2007kj},
the other parameter in the 
vertex $\Gamma_\mu$ is a momentum scale $\Lambda = 1.8$~GeV.
In Ref.~\cite{Haberzettl:2007kj}, there were also two additional 
parameters associated with the vertex $\Gamma_{\mu \nu}$.
They were adjusted to fit the inclusive cross sections for 
direct $J/\psi$ production at transverse momenta up to 10~GeV
that were measured by the CDF Collaboration.
Following Lansberg et al.,
we set the center-of-mass energy of the colliding $p$ and $\bar p$
to $\sqrt{s}=1.8$~TeV, we impose a pseudo-rapidity cut $|\eta|<0.6$ 
on the $J/\psi$, and we use the  Martin-Roberts-Stirling-Thorne (MRST) 
parton distribution function parton distributions
at leading order~\cite{Martin:2002dr}.
We take both the renormalization and factorization scales to be 
 $\sqrt{4 m_c^2+p_T^2}$. 
In Fig.~\ref{comp_predictions}, our result for the 
charm-pair rescattering ($s$--channel cut) contribution
is compared to the results from Refs.~\cite{Lansberg:2005pc}
and ~\cite{Haberzettl:2007kj}.
Our result is larger than that from Ref.~\cite{Lansberg:2005pc} 
by a factor that increases from about 2 at $p_T = 4$~GeV
to about 4 at $p_T = 15$~GeV.  Our result is more than two orders
of magnitude smaller than that from Ref.~\cite{Haberzettl:2007kj}.
Since the primary difference between Refs.~\cite{Lansberg:2005pc}  
and~\cite{Haberzettl:2007kj} is the model for the vertex $\Gamma_{\mu \nu}$,
our results suggest that the model for the vertex $\Gamma_{\mu \nu}$ used in
Ref.~\cite{Haberzettl:2007kj} gives an unrealistically large
contribution from the s-channel $c \bar c$ cut mechanism.

\begin{figure}[t]
\center
\includegraphics[scale=0.9]{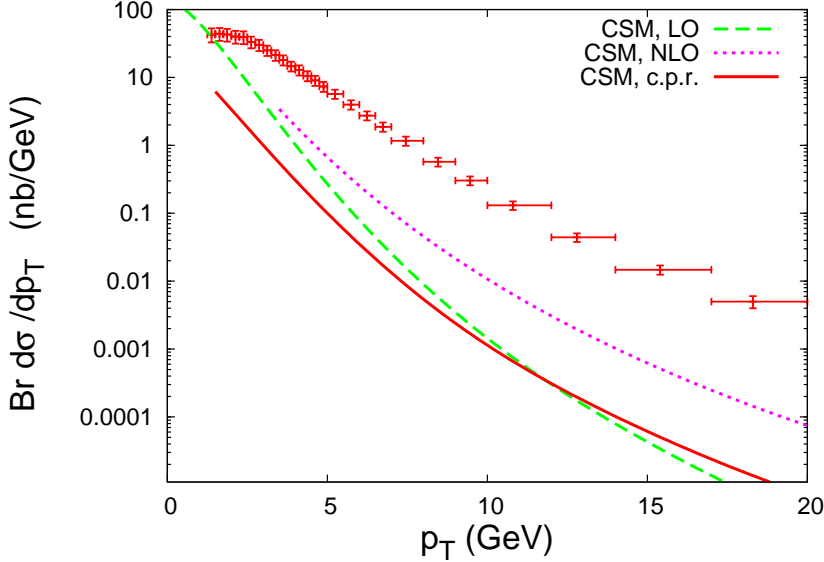}
\caption{Color-singlet contributions to the
differential cross section for direct $J/\psi$ production 
at the Tevatron compared to data from the CDF 
Collaboration~\cite{Acosta:2004yw,Abe:1997yz}.
The differential cross section $d \sigma/dp_T$ 
integrated over the rapidity interval $|y| < 0.6$ and
multiplied by the branching fraction for
$J/\psi \rightarrow \mu^+ \mu^-$ is shown as a function of $p_T$.
The curves are the color-singlet model terms
at LO (dashed line), through NLO (dotted line),
and at NNLO from charm-pair rescattering (solid line).}
\label{comp_with_data}
\end{figure}

We now proceed to assess the phenomenological importance 
of charm-pair rescattering by comparing it to experimental data 
and to other CSM contributions.  The data on inclusive $J/\psi$
production at the Tevatron published by the CDF collaboration 
are shown in Fig~\ref{comp_with_data}. 
The differential cross section 
for prompt $J/\psi$ measured in Ref.~\cite{Acosta:2004yw}, which
includes a factor Br$=0.0588$ for the branching fraction 
of $J/\psi$ into $\mu^+ \mu^-$, has been converted
into a differential cross section for direct $J/\psi$ by
multiplying by the fraction of $J/\psi$'s that are produced directly, 
which was measured in Ref.~\cite{Abe:1997yz}. 
For a consistent comparison with these data, we set 
$\sqrt{s}=1.96$ TeV and we impose a rapidity cut $|y|<0.6$
in our calculations of CSM contributions.
We set $m_c=1.5$ GeV, which is natural as it is 
close to half the mass of the $J/\psi$.
We set $\langle \mathcal O_1^{J/\psi}(^3S_1) \rangle=1.16$~GeV$^3$,
which is the value obtained from the wave function at the
origin in the Buchm\"uller-Tye potential~\cite{Eichten:1995ch}.
We also use a more recent set of parton distributions~\cite{Pumplin:2002vw}.
We choose the factorization and renormalization scales 
to be $\sqrt{4 m_c^2+p_T^2}$.
In Fig.~\ref{comp_with_data}, the NNLO contribution from charm-pair 
rescattering to the CSM term in the differential cross section 
is compared to the LO and NLO terms.
The overall charm-pair rescattering contribution is a subdominant fraction 
of the CSM term. At $p_T = 4$~GeV, it 
 is smaller than the LO term 
by about a factor of 3. As it decreases more slowly  
with $p_T$, it is larger than the LO term for $p_T > 12$ GeV.
This is expected since the differential cross section
associated to the leading order CSM production mechanism
shows a well-known strong kinematic suppression  at large $p_T$.
Compared to the NLO term, the charm-pair rescattering contribution
is negligible in the whole $p_T$ range accessible at the Tevatron:
it is suppressed by a factor that increases from about 7 at $p_T = 4$~GeV
to about 10 at $p_T = 20$~GeV.  
We also note that the CSM yield at NLO 
substantially underestimates the Tevatron data.
We conclude that charm-pair rescattering can account 
only for a small fraction of the cross section 
measured by the CDF collaboration.

\section{Summary}

We have identified the $s$--channel cut mechanism for quarkonium production
proposed by Lansberg, Cudell, and Kalinovsky~\cite{Lansberg:2005pc}
as $Q \bar Q$ rescattering.  In the NRQCD factorization approach, 
this mechanism gives a  contribution to the color-singlet model (CSM) term 
at NNLO and it can be calculated without introducing any additional
phenomenological parameters beyond the usual NRQCD matrix elements.
In the case of $J/\psi$ production at the Tevatron, the NNLO 
contribution from charm-pair rescattering is comparable to the LO
CSM term but is much smaller than the NLO term.
The charm-pair rescattering contribution is substantially smaller 
than the CDF data in the whole $p_T$ range accessible at the Tevatron.
We conclude that charm-pair rescattering (the $s$--channel cut mechanism) 
is not a dominant mechanism
for charmonium production in high-energy collisions.

\begin{acknowledgments}
P.A.\ was supported by the Fonds National de la Recherche
Scientifique and by the Belgian Federal Office for Scientific, 
Technical and Cultural Affairs through the 
Interuniversity Attraction Pole No.~P6/11.
E.B.\ was supported in part by the Department of Energy
under grant DE-FG02-91-ER40690 
and by the Alexander von Humboldt Foundation.
\end{acknowledgments}


\end{document}